# Incoherent transport across the strange metal regime of overdoped cuprates

J. Ayres*[1,2], M. Berben*[2], M. Čulo[2], Y.-T. Hsu[2], E. van Heumen[3], Y. Huang[3], J. Zaanen[4], T. Kondo[5], T. Takeuchi[6], J. R. Cooper[7], C. Putzke[1], S. Friedemann[1], A. Carrington[1], and N. E. Hussey[1,2]

[1]H. H. Wills Physics Laboratory, University of Bristol, Tyndall Avenue, Bristol BS8 1TL, United Kingdom

[2]High Field Magnet Laboratory (HFML-EMFL) and Institute for Molecules and Materials, Radboud University, Toernooiveld 7, 6525 ED Nijmegen, Netherlands

[3]Van der Waals-Zeeman Institute, University of Amsterdam, Postbus 94485, 1090 GL Amsterdam, Netherlands

[4]The Institute Lorentz for Theoretical Physics, Leiden University, PO Box 9506, 2300 RA Leiden, Netherlands

[5]Institute for Solid State Physics, University of Tokyo, Kashiwa-no-ha, Kashiwa, Japan

[6]Toyota Technological Institute, Nagoya 468-8511, Japan

[7]Department of Physics, University of Cambridge, Madingley Road, Cambridge, CB3 0HE, United Kingdom

*These authors contributed equally to this work

**Strange metals possess highly unconventional electrical properties, such as a linear-in-temperature (*T*) resistivity,[1-6] an inverse Hall angle[7-9] that varies as $T^2$ and a linear-in-field (*H*) magnetoresistance.[10-13] Identifying the origin of these collective anomalies has proved profoundly challenging, even in materials such as the hole-doped cuprates that possess a simple band structure. The prevailing dogma is that strange metallicity in the cuprates is tied to a quantum critical point at a doping *p*\* inside the superconducting dome.[14,15] Here, we study the high-field in-plane magnetoresistance of two superconducting cuprate families at doping levels beyond *p*\*. At all dopings, the magnetoresistance exhibits quadrature scaling and becomes linear at high *H*/*T* ratios, indicating that the strange metal regime extends well beyond *p*\*. Moreover, its magnitude is found to be much larger than predicted by conventional theory and is insensitive to both impurity scattering and magnetic field orientation. These observations, coupled with analysis of the zero-field and Hall resistivities, suggest that despite having a single band, the cuprate strange metal region hosts two charge sectors, one containing coherent quasiparticles, the other scale-invariant `Planckian' dissipators.**

Our understanding of metallic behaviour is rooted in the concept of coherent quasiparticles, low-lying excitations that propagate through a periodic lattice as Bloch waves. Their interactions with all other quasiparticles are encapsulated in a description that subsumes all many-body interactions into a small number of renormalisation parameters, including a renormalised effective mass *m*\*. This treatment, enshrined in Fermi-liquid (FL) theory, also laid the foundations for the BCS theory of superconductivity. The condition for quasiparticle coherence is met once its decay rate Γ becomes smaller than its excitation energy *ε* which, in a FL, is guaranteed by the relation Γ ~ $\varepsilon^2$. In correlated metals, electron interactions can become so strong that the quasiparticle description may no longer be valid, even though the resistivity *ρ*(*T*) retains metallic character. This class of materials includes so-called bad metals, where the quasiparticle concept breaks down at high $T$,[16] and strange metals, where it breaks down even at low $T$.[3] In the latter, quasiparticle decoherence is implicit in the linear dependence of $\Gamma$ on $T$ and $\varepsilon$, as well as in its associated 'Planckian' timescale $\tau_\hbar$ (= $\hbar/\Gamma = \hbar/ak_BT$) where $a$ is of order unity[3,5,17]) that is conjectured to be the shortest time in which energy can be dissipated.[18-20]

The high-$T_c$ cuprates are exceptional in that they exhibit both bad and strange metallic behaviour; the in-plane resistivity $\rho_{ab}(T)$ near optimal doping grows linearly in $T$ right up to its melting point with a slope defined by the Planckian time. Such a condition is believed to occur in the strongly interacting critical state anchored at a quantum critical point (QCP) where a phase transition is tuned to zero temperature. Within the quantum critical 'fan' above the QCP, $\rho(T)$ typically displays the same $T$-linear behaviour seen in cuprates.[2,4,6] While some evidence exists for a QCP in certain cuprate materials near a hole doping $p^*$ = 0.19[15] – where the normal state pseudogap vanishes – recent analysis of the anti-nodal states across $p^*$ in photoemission,[21] as well as analysis of transport and thermodynamic data,[22,23] has cast doubt on the ubiquity of a QCP at $p^*$. In particular, there is as yet no evidence for a phase transition below $p^*$ that is suppressed to $T$ = 0 as $p$ is increased.[23]

In this report, we explore further the issue of criticality in cuprates via a study of the in-plane magnetoresistance (MR) of two families of heavily overoped (OD), single layer cuprates – (Pb/La)-doped $Bi_2Sr_2CuO_{6+\delta}$ (Bi2201) and $Tl_2Ba_2CuO_{6+\delta}$ (Tl2201) – across an extended region of the phase diagram as indicated in Figure 1a (see Methods and Extended Data Fig. 1 for more details on the samples themselves). It is widely assumed that beyond $p^*$, conventional FL physics (e.g. $\rho_{ab}(T) \sim T^2$) is re-established.[14] Yet across this entire regime, the low-$T$ resistivity retains a finite $T$-linear component.[3,5,24] Although the 'strange' component $\rho_{ab}(T) \sim AT$ (Figure 1b) gradually diminishes as a function of $p$,[3,24] its persistence is difficult to explain within the usual QCP scenarios.[3] Moreover, as shown in Figure 1c, recent high-field Hall effect studies have revealed an anomalous drop in the Hall number[25] from 1 + $p$ to $p$[26] over a similar doping range, while the observed decrease in superfluid density with overdoping is claimed to be at odds with BCS theory.[27] Hence, descriptions of superconducting OD cuprates as conventional metals seem to fail in capturing the full experimental picture. Here, we reveal that the in-plane MR is also highly unconventional.

**Magnetoresistance scaling**

In a FL (Drude) metal, the magneto-transport is determined uniquely by the magnetic field strength $H$, which enters via the cyclotron frequency $\omega_c = e\mu_0H/m^*$ and the transport relaxation time $\tau_{tr}$. Both quantities then combine into a dimensionless parameter $x = \omega_c\tau_{tr}$. The low-field Hall angle $\tan\theta_H \propto x$ while typically, the longitudinal MR $\Delta\rho(H)/\rho(0) = [\rho(H) - \rho(H = 0)]/\rho(H = 0) \propto x^2 \propto (H/\rho(0))^2$. The latter relationship ($\Delta\rho(H)/\rho(0) \propto (H/\rho(0))^2$) is known as Kohler's scaling and is found in many standard metals. While various scenarios can lead to a violation of Kohler's scaling, such as $T$-dependent anisotropic scattering or multi-carrier systems where the mobility of each carrier-type has a distinct *T*-dependence,[28] the violation reported recently in the quantum critical metal BaFe$_2$(As$_{1-x}$P$_x$)$_2$ (P-Ba122)[10] suggests an entirely new form of MR scaling. At the antiferromagnetic QCP in P-Ba122, $\Delta\rho(H,T) = \rho(H,T) - \rho(0,0) = \sqrt{(\alpha k_B T)^2 + (\gamma \mu_B \mu_0 H)^2}$ where *γ* and *α* are constants independent of *T* and *H* and *γ*/*α* ~ 1. By analogy with the Drude metal, we can re-express this MR response in terms of a new dimensionless parameter $x_\hbar = \beta \mu_0 H/T$, where *β* = *γμ*$_B$/*αk*$_B$ and thus $\Delta\rho(T,x) = \alpha k_B T \sqrt{1 + x_\hbar^2}$. This implies that the timescale associated with the field (1/$\omega_\hbar$) plays a similar role to the thermal time $\tau_\hbar$ (= $\hbar/k_B T$) in this state (though we stress here that $\omega_\hbar$ is not necessarily associated with cyclotron motion). Starting from generalities of thermal quantum field theory, it is unclear why this should be the case,[29] and even within a more conventional effective medium approach, such 'Planckian quadrature' behaviour requires significant fine tuning of parameters.[30,31] The fact that similar behaviour has now been reported in both the electron-doped cuprate La$_{2-x}$Ce$_x$CuO$_4$ (LCCO)[11] and in FeSe$_{1-x}$S$_x$[13] at or near their putative QCPs suggests that it is a generic feature of quantum critical metals.

In Tl2201 and Bi2201, at all doping levels marked in Figure 1a, the MR is found to exhibit a similar crossover from $H^2$ to *H*-linear dependence with increasing field strength – as exemplified in Fig. 1d for a highly OD Bi2201 sample ($T_c \leq$ 1 K) at *T* = 4.2 K and in Fig. 2A for a OD Tl2201 sample ($T_c \leq$ 26.5 K) over a wide temperature range. In contrast to P-Ba122, however, $\rho_{ab}(H,T)$ does not collapse onto a single line when plotted as $\Delta\rho_{ab}(H,T)/T$ vs. *H*/*T*. In OD cuprates, this is to be expected since, in the absence of a magnetic field, $\rho_{ab}(T)$ has a super-linear (not *T*-linear) dependence. Hence, the quadrature expression alone is not sufficient to describe $\rho_{ab}(H,T)$ completely. However and remarkably, the derivatives d$\rho_{ab}(H,T)$/d*H* for all our Tl2201 and Bi2201 samples are found to collapse onto a universal curve when plotted against *H*/*T* (see Fig. 2c-j). Moreover, the form of the derivative is found to be identical to that for a pure quadrature MR, indicating that all terms in $\rho_{ab}(H,T)$ that are dependent upon both field *and* temperature can be well described by the quadrature expression. A more complete (and still general) form of $\rho_{ab}(H,T)$ is therefore:

$$\rho_{ab}(H,T) = \mathcal{F}(T) + \sqrt{(\alpha k_B T)^2 + (\gamma \mu_B \mu_0 H)^2} \tag{1}$$

where $\mathcal{F}(T)$ is an additional term in the zero-field resistivity that accounts for the super-linear form of $\rho_{ab}(T)$ but does not of itself display significant MR. Taking the derivative of the raw $\rho_{ab}(H,T)$ data thus isolates the quadrature MR and reveals the hidden *H*/*T* scaling (see Extended Data Fig. 2 for more details of how the *H*/*T* scaling is revealed).

Panels c-j of Fig. 2 highlight the data collapse observed in all samples and at all temperatures studied, indicating that cuprates exhibit anomalous scale-invariant MR across an extended region of the strange metal regime. In another hole-doped cuprate La$_{2-x}$Sr$_x$CuO$_4$ (LSCO) near *p*\* = 0.19, the MR is also found to become *H*-linear at high fields[12] (green diamond in Figure 1a). In that report, it was concluded that there were in fact two quantum critical fans in cuprates, one in the $T - p$ plane and one in the $H - p$ plane, both of which terminate at a QCP at $p^*$. Our observation of a similar MR response at low $H/T$ at $p \gg p^*$ (for a recent discussion of the location of $p^*$ in Tl2201 and Bi2201, see ref. [25]) reveals that, just as the $T$-linear $\rho_{ab}(T)$ persists over a wide doping range beyond $p^*$,[3] so too does the anomalous linear MR. A similar extended region of $T$-linear resistivity and $H$-linear MR is also observed in electron-doped LCCO, though there, the MR does not follow precisely the quadrature form,[11] the reasons for which are unclear. In LSCO, departures from quadrature behaviour are also observed,[31] possibly due to the presence of the pseudogap, though clearly more measurements across $p^*$ would be required to establish the role of the pseudogap in causing a breakdown in $H/T$ scaling.

Finally, as shown in panels k and l of Fig. 2, while there is a marked difference in the relative size of $\beta$ (= $\gamma\mu_B/\alpha k_B$) in Tl2201 and Bi2201, neither family exhibits any significant $p$-dependence in $\beta$. Since $\gamma$ is found to be comparable in both families, the larger values of $\beta$ in Tl2201 are likely to be related to the smaller $T$-linear coefficient of the zero-field resistivity in Tl2201, shown in Fig. 1b (we note though that the coefficients of the $T^2$ term in both families are comparable[29]). The ratio $\gamma/\alpha$ determines the field scale at which there is a crossover from quadratic to linear MR. A lack of doping dependence in $\beta$ implies that there is no quantum critical 'fan'. Specifically, there is no indication that the $H$-linear behaviour only extends to lower magnetic fields (at a given temperature) upon approach to $p^*$, as implied for LSCO.[12] Moreover, the fact that $\gamma \sim \alpha$ (panels k and l) indicates that the quadrature MR has the same origin as the $T$-linear resistivity, implying that models based on real-space inhomogeneity[31,32] are not applicable here. This is reaffirmed by the observed MR responses of Tl2201 and Bi2201 being so similar despite them having very different levels of electronic inhomogeneity.[33,34]

**Signatures of incoherent transport**

Several features of the uncovered MR response are surprising and reveal new aspects of the strange metal phase in hole-doped cuprates. In Fig. 3a-d, we show evidence that the magnitude of the quadrature MR in OD cuprates is far greater than one would expect from standard Boltzmann theory and is insensitive to $1/\tau_0$, the impurity scattering rate. The dashed lines in panels a and b represent estimates for $\rho_{ab}(H)$ at $T = 0$ K for both Bi2201 and Tl2201 (using the known Fermi surface parameters – see Methods for details). In both cases, the observed MR is two orders of magnitude larger. In a FL, since $\Delta\rho(H)/\rho(0) \propto (\omega_c\tau_{tr})^2$, where $1/\tau_{tr} = 1/\tau_0 + 1/\tau_{in}$ (with $1/\tau_{in}$ the inelastic scattering rate), the size of the MR is extremely sensitive to the residual resistivity $\rho_0$. Panels c and d of Figure 3, however, show that the magnitude of the linear slope of the transverse MR at high $H$ and low $T$ is very similar in the two families, despite the fact that $\rho_0$(Bi2201) ~ 10 x $\rho_0$(Tl2201) (and hence $1/\tau_0$(Bi2201) ~ 10 x $1/\tau_0$(Tl2201) which according to Kohler's rule would lead to a large suppression of the MR in Bi2201). This indicates that the fundamental timescale associated with the Planckian quadrature MR is largely insensitive to elastic scattering.

The next striking feature in the data is the lack of anisotropy in the MR response. As shown in Figure 3e/f, the longitudinal MR (with **H** ∥ *I* ∥ *ab*) for both families exhibits the same quadrature form as the transverse MR (with **H** ∥ *c*) (shown in panels c and d), with a similar magnitude. In contrast, the quadrature MR in Ba-122 was found to be anisotropic and tied to the crystallographic lattice rather than to the current direction.[35] These findings may point more to a MR response driven by Zeeman (i.e. spin) physics, though clearly, further investigations will be needed to confirm this (see also Methods for an expanded discussion of this point). Whatever its origin, it follows directly from experiment that this Planckian quadrature MR is highly anomalous; the scaling itself is not at all understood, it appears to be independent of $\rho_0$ with an isotropic MR that has no intrinsic Hall response of its own (see below). These signatures of a possible non-orbital origin, coupled with its intimate association with the *T*-linear resistivity, suggests that the quadrature MR is itself a consequence of incoherent, non-quasiparticle transport persisting down to low *T*.

**Dual character of the strange metal**

This brings us to arguably the most profound finding. The insensitivity of the in-plane MR to field orientation contrasts markedly with the strongly angle-dependent interlayer magnetoresistance (ADMR) found in OD Tl2201 (*p* > 0.27, $T_c$ < 30 K) and modelled previously using Boltzmann transport theory to map out its entire Fermi surface.[36] Importantly, the Fermi surface derived from ADMR was found to agree with that determined both by ARPES[37] and by quantum oscillations[33] for these highly doped crystals. Subsequent *T*-dependent ADMR studies[38] revealed a momentum-dependent scattering rate $1/\tau(k)$ in heavily OD Tl2201 that could self-consistently explain both the *T*- and *H*-dependence of the in-plane Hall resistivity.[25,38] Thus, the field and temperature dependence of the Hall response appears to be well described by conventional Boltzmann theory. As shown in Extended Data Figs. 4 and 5, it was not possible to replicate the in-plane MR response using the same parameterisation, nor any other parameterisation (for $\tau(k, T)$ and the Fermi velocity $v_F(k)$) based on the same Fermi surface geometry. In particular, the fact that the measured MR is both insensitive to the magnitude of impurity scattering rate (panels c-f of Fig. 3) and at least an order of magnitude larger than the anticipated Boltzmann response (at low *T* - dashed line in Fig. 3a,b) dictates that no variant of the Boltzmann formalism can reproduce the experimental data. This dichotomy hints strongly at the presence of two distinct contributions to the in-plane transport in OD cuprates, one described by conventional transport theory (albeit with anisotropic scattering), the other highly non-FL but characteristic of Planckian dissipation physics. We reiterate here that these highly OD cuprates appear to have no direct relation to a QCP. A similar coexistence of two charge sectors was also deduced recently from high-field magneto-transport studies on both FeSe$_{1-x}$S$_x$ (where they were assumed to add in parallel)[13] and P-Ba122 (where they were assumed to add in series).[39] Both these systems possess multiple pockets of charge carriers which might provide a natural framework for the appearance of two sectors. In OD cuprates, on the other hand, there is only one Fermi sheet. Moreover, the observation of quantum oscillations (QO) in OD Tl2201 consistent with orbits encircling the whole sheet would seem to rule out incoherence on any significant portion of the Fermi surface, at least for *p* > 0.27. It is noted, however, that under certain circumstances, QO are observable even in insulators[40] which by their nature do not possess a conventional Fermi surface. Clearly, further theoretical work will be required to establish if all the now known characteristics of the strange metallic state can be explained in a single framework; one that also includes the emergence of the high temperature superconducting state.

**Acknowledgements:** The authors would like to thank M. Allan, J. G. Analytis, I. Božovič, M. S. Golden, B. Goutéraux, C. Pépin, K. Schalm, H. Stoof and S. Vandoren for insightful discussions during the course of this work. We also thank S. Smit and L. Bawden for initial characterization of some of the Bi2201 single crystals and L. Malone for assistance in the growth of the Tl2201 single crystals. JA acknowledges the support of the EPSRC-funded CMP-CDT (Ref. EP/L015544/1) and an EPSRC Doctoral Prize Fellowship (Ref. EP/T517872/1). AC acknowledges support of the EPSRC (Ref. EP/R011141/1). We also acknowledge the support of the High Field Magnet Laboratory (HFML) at Radboud University (RU), member of the European Magnetic Field Laboratory (EMFL – also supported by the EPSRC, Ref. EP/N01085X/1), and the former Foundation for Fundamental Research on Matter (FOM), which is financially supported by the Netherlands Organisation for Scientific Research (NWO) (Grant No. 16METL01), 'Strange Metals'. Finally, part of this work was supported by the European Research Council (ERC) under the European Union's Horizon 2020 research and innovation programme (Grant Agreements no. 835279-Catch-22 and 715262-HPSuper).

**Author statements:** JA, MB, SF and AC and NEH conceived the overall project. JA, MB, MČ, YTH, CP and NEH performed the high-field measurements. YH, EvH, JRC, CP, TK and TT grew and characterized the single crystal samples. JA and AC performed the SCTIF calculations. JA, MB, JZ and NEH wrote the manuscript with input from all of the co-authors.

**Figure Captions**

**Figure 1. The strange metal regime of overdoped cuprates. a:** Temperature $T$ vs. doping $p$ phase diagram showing the superconducting ($T_c$ vs. $p$) dome (dotted lines) for the single-layer hole-doped cuprates $Tl_2Ba_2CuO_{6+\delta}$ (Tl2201), La/Pb-doped $Bi_2Sr_2CuO_{6+\delta}$ (Bi2201) and $La_{2-x}Sr_xCuO_4$ (LSCO).[12] The thick orange dashed line marks (approximately) the temperature onset $T^*$ for physical manifestations of the opening of the normal state pseudogap in the single-particle excitation spectrum and the temperature at which the resistivity deviates from its high-$T$ $T$-linear dependence. The faint dashed line on the right side of the graph indicates the temperature below which the resistivity in LSCO is purely quadratic. The red squares, blue circles and green diamonds indicate doping levels at which the in-plane magnetoresistance (MR) is found to vary linearly with magnetic field (at high field strengths) in Tl2201, Bi2201 (this work) and LSCO[12] respectively. **b**: Coefficient $A$ of the low-$T$ $T$-linear resistivity in Tl2201 (red squares),[24] Bi2201 (blue circles)[25] and LSCO (green diamonds),[3] normalized by the interlayer distance $d$. Note the different ordinate axes to accommodate the (smaller) $A$ values for Tl2201. **c:** Evolution of the low-$T$ Hall number $n_H(0)$ across the strange metal regime in Tl2201 (red squares) and Bi2201 (blue circles), as determined from Hall resistivity measurements in high magnetic fields.[25] A crossover from $n_H(0) \sim p$ to $n_H(0) \sim 1 + p$ is found to occur across a wide doping range beyond $p^*$, the doping level at which the pseudogap vanishes. The grey dashed line is a guide to the eye. At low doping in LSCO, $n_H(0)$ follows closely the number of doped holes,[26] as indicated by the green diamonds. The evolution of $n_H(0)$ in LSCO beyond $p = 0.08$ is difficult to obtain from Hall effect measurements due to the onset of charge order and a change in the Fermi surface geometry around $p = 0.20$, when the Fermi level crosses the van Hove singularity. **d:** Transverse in-plane MR of a heavily overdoped Bi2201 sample ($T_c < 1$ K) at $T = 4.2$ K, showing the crossover from quadratic MR at low-field to $H$-linear MR at higher field as indicated by dashed lines. The error bars in panels **a**, **b** and **c** are reproduced from published data and reflect uncertainty in doping level, where known, as well as geometrical uncertainty in the sample dimensions and the positioning of the voltage contacts. Since transport is a one-dimensional probe of superconductivity, we ascribe an error margin of $p = 0.005$ to the doping levels of Tl2201 and Bi2201 defined by their $T_c$ values, except for the 0K Bi2201 sample, whose error margin is set at $p = 0.01$ in panel **a**.

**Figure 2. Quadrature scaling of the in-plane magnetoresistance (MR) in heavily overdoped cuprates. a:** In-plane transverse (**H** ∥ $c$) MR $\rho_{ab}(H)$ of OD Tl2201 ($T_c = 26.5$ K) up to 35 T for a temperature range between 1.4 K and 60 K. **b:** The derivatives with respect to the magnetic field for the MR curves shown in panel **a**. (In panel **i**, the same data are shown plotted against $H/T$). **c-j:** Scaled derivatives of the in-plane MR at different fixed temperatures between 4.2 K and 60 K for various Tl2201 and Bi2201 samples with $T_c$ values as indicated. To emphasize the similarity of the MR response of all the measured samples, the $y$-axis has been multiplied by $1/\mu_B \gamma$ and the $x$-axis by $\beta = \gamma \mu_B / \alpha k_B$) (see Eq. (1) for definitions of $\alpha$ and $\gamma$). In this way, all the data collapse onto the same form given by the derivative of the function $y = \sqrt{1 + x^2}$ as indicated by dashed lines. Note that the sections of individual curves that reside within the mixed state are here plotted faintly, since only in the normal state can the quadrature MR be probed. **k,l:** Plots of the scaling parameter $\beta$ for Bi2201 and Tl2201 respectively plotted versus doping (top axis) and $T_c$ (bottom axis). For both materials, $\beta$ is found to be independent of doping (and $T_c$) as indicated by the horizontal dashed lines. The error bars in panels **k,l** reflect the sensitivity of $\beta$ to details of the fitting procedure, in particular the field and temperature range over which the fits are performed.

**Figure 3. Evidence for incoherent transport in heavily overdoped cuprates. a-b:** In-plane transverse (**H** ∥ $c$) MR $\rho_{ab}(H)$ of OD Tl2201 ($T_c = 26.5$ K) and OD Bi2201 ($T_c = 7$ K) up to 35 T. Note the much larger residual resistivity $\rho_0$ in the Bi2201 crystal. For ease of comparison, the $y$-axes of both panels span the same range in absolute resistivity. The dashed lines in both panels are estimates of the orbital transverse MR at $T = 0$ (see Supplementary Material for details). **c-f:** Derivatives $d\rho_{ab}/dH$ for the same single crystals with **c/d: H** ∥ $c$ (transverse MR) and **e/f: H** ∥ $ab$ (longitudinal MR). The form and magnitude of the MR is comparable for all materials and field orientations.

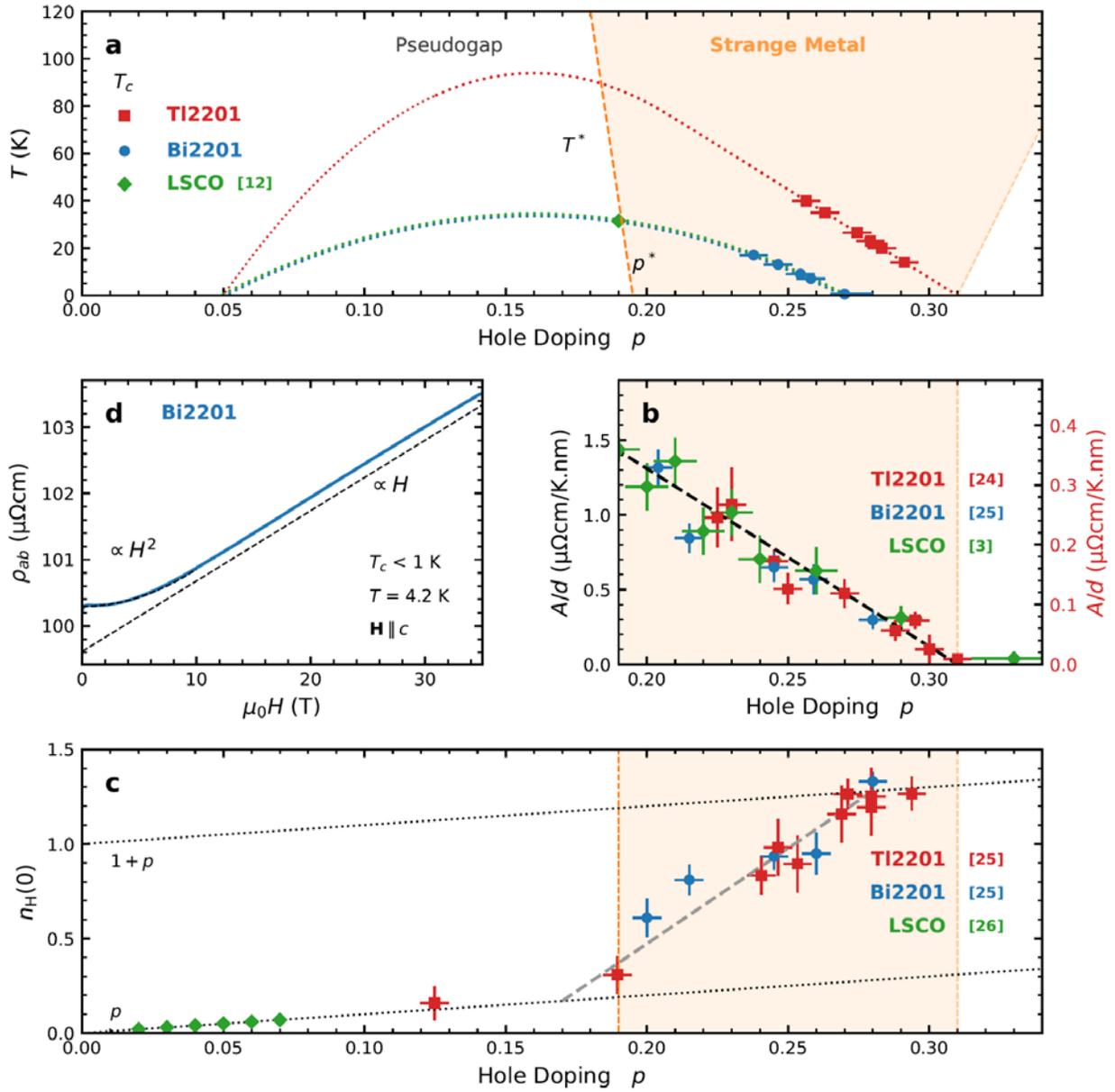

Figure 1

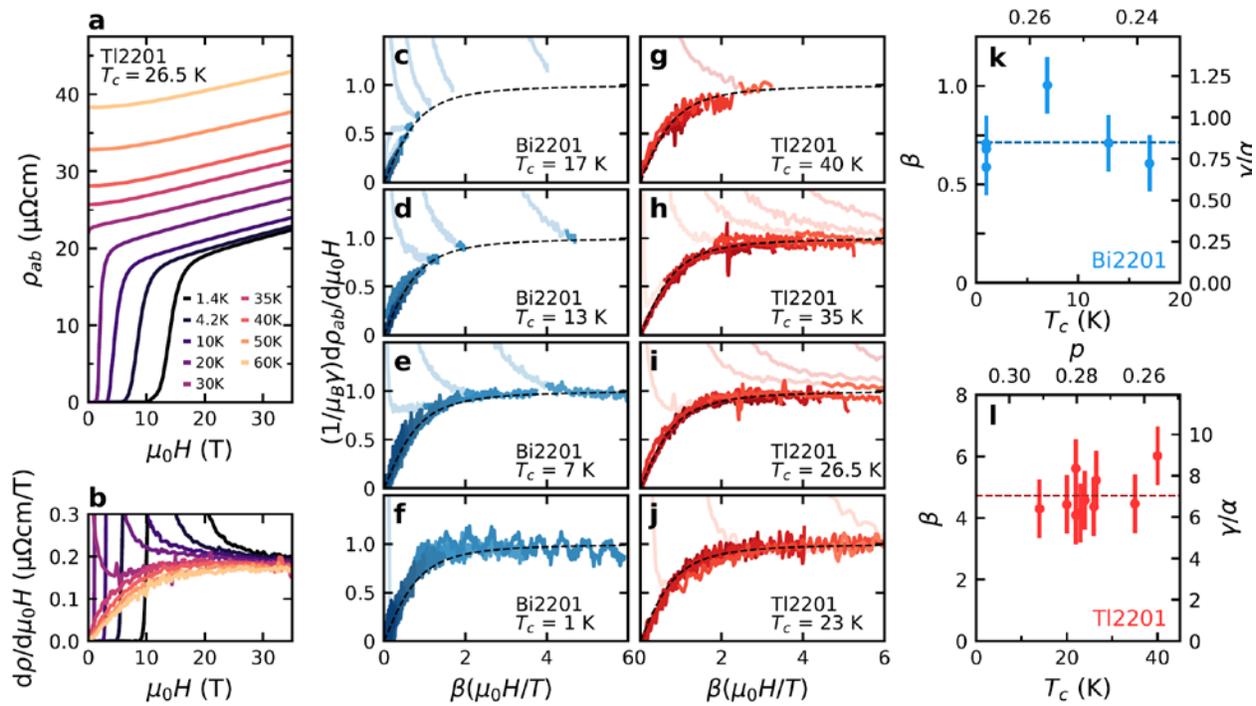

Figure 2

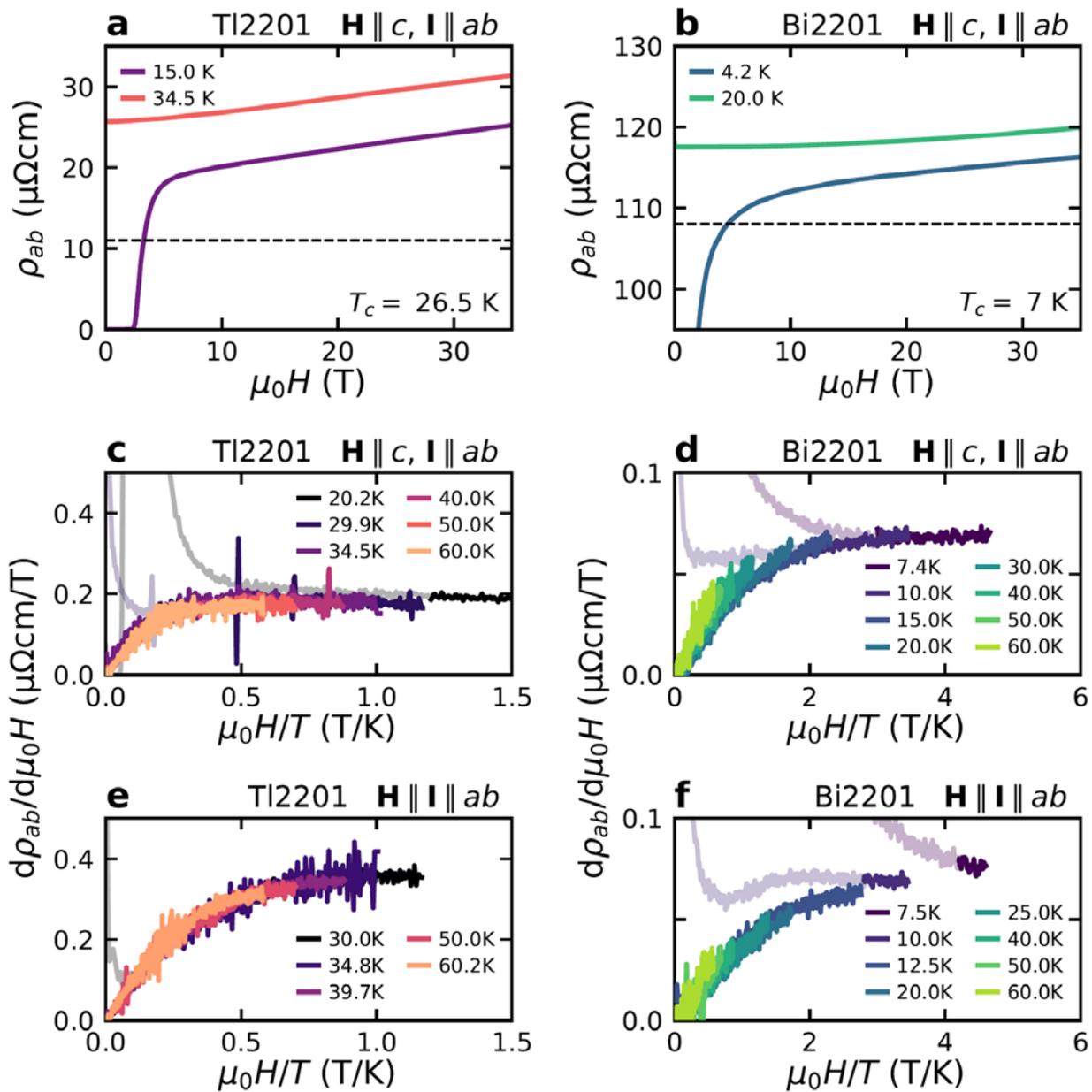

Figure 3

# Methods

## Sample preparation and measurement

Single crystals of Tl2201 were grown via a self-flux method similar to that detailed in ref. [41] and subsequently annealed in flowing oxygen at elevated temperatures to set their oxygen content (and hole doping). The usual parabolic relationship[42] between doping $p$ and $T_c$ fails to account for the higher range of dopings over which superconductivity persists in Tl2201 as determined through quantum oscillation studies.[33] Instead, a simple linear parameterisation $p = (228.4 - T_c)/737.2$ was used which is found to more closely match $T_c(p)$ from the aforementioned quantum oscillation studies at high dopings.

Electrical contacts were made by sputtering gold pads with a copper sub-layer onto both the top and sides of a sample before attaching gold wires with Dupont 4929 silver paint. The contacts were then annealed in flowing oxygen at 450 °C for 15 minutes prior to annealing to set $T_c$. Typical contact resistances were around 1 Ω. Care was taken to ensure that the sputtered gold and silver paint covered the sides of the sample in order to minimise contributions to the measured MR from currents flowing along the $c$-axis. Typical in-plane sample dimensions were 800 μm x 250 μm with a thickness of 10-15 μm.

Single crystals of La/Pb doped Bi2201 were grown using the floating zone technique at two different sites. The doping was estimated from the measured $T_c$ using the parabolic relation[42]: $1 - (T_c/T_c^{max}) = 82.6(p - 0.16)^2$ with $T_c^{max}$ = 35 K. Electrical contacts were made to bar-shaped samples cut from the as-grown crystals by attaching gold wires with Dupont 6838 silver paint. The contacts were then annealed in flowing $O_2$ at 450 °C for 10 minutes. Unlike Tl2201, these anneal times are sufficiently short to have a negligible effect on $T_c$. Typical contact resistances were around 1 Ω. Typical sample dimensions, meanwhile were 1000 μm x 250 μm with thicknesses varying between 6 and 25 μm.

A standard four-point ac lock-in detection method was used to measure the in-plane resistivity of all samples. Extended Data Tables 1 and 2 list the contacted samples used in this study and Extended Data Figure 1 shows a representative set of $\rho_{ab}(T)$ curves for samples with different $T_c$ values. Note that all of the samples exhibit the super-linear $T$-dependence characteristic of OD cuprates with doping levels beyond $p^*$. This super-linear behaviour in the zero-field resistivity dictates that pure $H/T$ scaling of the MR can never be realised in these samples (see following section). Measurements in magnetic fields up to 35 T were performed at the HFML in Nijmegen. The field was oriented either parallel or perpendicular to the $CuO_2$ planes using a rotating sample stage.

The lack of agreement with standard Boltzmann theory in the MR response highlighted below, coupled with the lack of any angle-dependence in the MR when the field is applied either perpendicular or parallel to the current, points towards a non-orbital origin for the in-plane MR of highly OD cuprates. High-$T_c$ cuprates, however, are layered compounds with a highly anisotropic electronic state characterised by electrical resistivity ratios $\rho_c/\rho_{ab}$ as high as $10^6$ for Bi2201.[43] Moreover, in OD Tl2201, the interlayer MR $\Delta\rho_c(H)$ has a quasi-linear dependence on $H$ at high fields (**H** || $c$)[44] due to the specific $c$-axis warping of the FS in (body-centred-tetragonal) Tl2201 that leads to an effective cancellation of the $c$-axis velocity around any in-plane cyclotron orbit.[45] Hence, it is important to eliminate $c$-axis mixing of the resistivity tensor as a source of the $H$-linear MR and quadrature scaling reported here.

As mentioned above, gold pads were sputtered onto both the top and sides of all samples in order to minimise such mixing. Importantly, the absolute magnitudes and $T$-dependencies of the resistivities shown in Extended Data Figure 1 are comparable with those reported in the literature.[41,46,47] Moreover, the variation in absolute resistivities across our series of Bi2201 samples is no greater than the geometrical uncertainty (± 20%). (The variation in Tl2201 is larger due to the fact that these samples were mounted for measurements inside a pressure cell and due to their small size, larger contact pads relative to the sample dimensions were required.) Due to the very large electrical resistivity anisotropy in both Bi2201 and Tl2201, any mixing of the current paths would lead to a marked increase in the absolute value of the as-measured resistivity relative to the intrinsic in-plane response as well as a different $T$-dependence. The lack of either aspect in our measurements is a strong indication that there is no $c$-axis mixing. Similarly, the fact that all samples exhibit the same MR, both in terms of its magnitude and field dependence, is incompatible with a scenario in which $c$-axis mixing was present, as its influence would be randomised between samples. Finally, we note that the anisotropic resistivity ratios in Bi2201 are approximately 3 orders of magnitude larger than in Tl2201, yet the quadrature MR is comparable in all samples studied.

## Quadrature scaling in overdoped cuprates

In BaFe$_2$(As$_{1-x}$P$_x$)$_2$ (P-Ba122) at the critical doping $x_c$ = 0.31,[10] the following ansatz was found to accurately describe $\rho(H,T)$:

$$\rho(H,T) - \rho(0,0) = \sqrt{(\alpha k_B T)^2 + (\gamma \mu_B \mu_0 H)^2} \quad (1)$$

That is, when the total MR is subtracted by a temperature- and field-independent residual resistivity, the remaining MR is of a purely quadrature form. As a result, plots of $\rho(H,T)$- $\rho(0,0)/T$ versus $H/T$ collapse onto a universal curve. While the in-plane MR of OD Tl2201 and Bi2201 exhibits a very similar crossover from $H^2$ (at low $H$) to $H$-linear (at high $H$), the form shown in Eq. (1) is not sufficient to describe the MR in OD cuprates. An example of this is shown in Extended Data Fig. 2 for an OD Tl2201 with $T_c$ = 26.5 K. The raw MR curves are reproduced in panel a. Panels b and c, meanwhile, show the same data plotted as Δ$\rho(H,T)/T$ vs. $H/T$ as done for P-Ba122.[10] Although a reasonable collapse of the data is found at low $T$ (Extended Data Fig. 2b), the data deviate significantly from the scaling form at higher temperatures (Extended Data Fig. 2c). This deviation comes about due to the fact that the zero-field resistivity has a super-linear, rather than strictly linear $T$-dependence. Hence, while any single field-sweep (at a fixed temperature) can be fitted using the same form of the MR as above, a global fit to $\rho(H,T)$ is not possible. It is therefore necessary to include additional $T$-dependent terms (in the zero-field resistivity) to account for this discrepancy. Indeed, as shown in panels d and e of Extended Data Fig. 2, if $\rho(0,T)$ is subtracted from $\rho(H,T)$, rather than $\rho(0,0)$, the full data set duly collapse onto a single line over the entire temperature range studied.

A similar collapse can also be achieved by plotting the derivative d$\rho$/d$H$ vs. $H/T$, as shown in Extended Data Fig. 2g (the non-scaled data are plotted in panel f). This procedure of plotting the derivative versus $H/T$ was repeated for all samples and the results are summarised in Figure 2 of the main text.

## Success of Boltzmann formalism in overdoped cuprates

In highly overdoped, superconducting Tl2201, many of its resistive properties - the form of the zero-field resistivity $\rho_{ab}(T)$, the interlayer angle-dependent magnetoresistance (ADMR) Δ$\rho_c(\phi,\vartheta,T,H)$, the $T$-dependent Hall coefficient $R_H(T)$ and the field-dependent Hall resistivity $\rho_{xy}(T,H)$ - have been effectively understood and modelled within a Boltzmann transport framework and the Shockley-Chambers tube-integral formalism (SCTIF) in particular. These collective successes are summarised in Extended Data Figure 3. Extended Data Fig. 3a shows a representative set of $c$-axis ADMR sweeps at various azimuthal angles (as a function of polar angle) at $T$ = 50 K for an OD Tl2201 sample with $T_c$ = 15 K.[48] The ADMR traces were fitted by generalising to the case where the product $v_F\tau$ varies around the FS. Detailed aspects of this FS parameterisation were later confirmed in a comprehensive quantum oscillation study.[49] A two-dimensional projection of the resultant FS is plotted in Extended Data Fig. 3b. In addition to a small four-fold anisotropy in $k_F$ (of order 5%), the primary ingredient necessary to model the ADMR was a decomposition of $\tau^{-1}$ into two additive components: an isotropic $\tau^{-1}_{iso}$ that grows as $T^2$ (solid black line in Extended Data Fig. 3c) and an anisotropic $\tau^{-1}_{ani}$ (solid red line in Extended Data Fig. 3c) that is negligible at zero temperature but grows predominantly linearly with $T$ (with a small $T^2$ component). More specifically, the in-plane geometry was defined by the FS wave vector $k_F(\phi) = k_{00} + k_{40} \cos(4\phi)$, a scattering lifetime (also with 4-fold symmetry): $\tau(\phi) = \tau_0/(1 + \lambda \cos(4\phi))$ and a cyclotron frequency that is assumed to vary linearly with field ($\omega_c = \omega_c^0 H$). The decomposition was then parameterised as follows: $(1 - \lambda)/\omega_c^0\tau(T) = A_{iso} + B_{iso}T^2$ and $2\lambda/\omega_c^0\tau(T) = A_{ani} + B_{ani}T + C_{ani}T^2$. The parameters extracted from this analysis[48] are listed in Extended Data Table 3.

Knowledge of $k_F(\phi)$ (which itself can be used to define $v_F(\phi)$), $\lambda$, $\omega_c^0$ and $\tau_0$ is all that is required, in principle, to calculate the corresponding in-plane transport properties. The solid line in Extended Data Fig. 3d represents the resultant low-$T$ zero-field resistivity $\rho_{ab}(T)$. The $T$-linear and $T^2$ components deduced from the ADMR combine to generate a $\rho_{ab}(T)$ curve with a super-linear $T$-dependence that matches well the experimental data (black dots in Extended Data Fig. 3**d**).

Despite OD Tl2201 having a single, cylindrical FS, $R_H(T,H)$ has a non-trivial temperature and field dependence. For a sample with $T_c$ ~ 25 K, for example, the low-field $R_H(T,H \to 0)$ grows by approximately 50% from its low-$T$ value up to ~ 120 K before subsequently falling to a value at $T$ = 300 K that is comparable to $R_H(0,H \to \infty)$). At the lowest temperatures, $R_H(0,H)$ is field-independent (beyond the vortex state),[25] suggesting that the system is effectively isotropic (or in the high-field limit which is not the case since $\omega_c\tau < 1$. In this regime, $R_H(0,H \to \infty)$ has a value consistent with the carrier density deduced from quantum oscillation studies, as expected. At intermediate temperatures, $R_H(T,H)$ is suppressed in field, asymptotically approaching its low-$T$ isotropic value.[25] All of these complexities have been successfully reproduced using the same ADMR-derived parameterisation,[25] as summarised in panels e-h in Extended Data Fig. 3. With increasing temperature, the linear growth of $\tau_{ani}$ leads to an increase in the in-plane anisotropy of the mean-free-path $\ell(\phi)$. As a result, $R_H(T, H \rightarrow 0)$ becomes enhanced. As $H$ increases, this anisotropy is progressively 'smeared out' as the quasiparticles traverse more of the FS. Eventually, at the highest fields, $R_H(T, H \rightarrow \infty)$ again approaches $R_H(0, H \rightarrow \infty)$ (the isotropic case). For $T > 125$ K, the system approaches the isotropic-$\tau$ limit, returning $R_H(T, H \rightarrow 0)$ to a value similar to that found at 0 K. The overall evolution of $R_H(T,H)$ thus mirrors the evolution of $\ell(\phi,T,H)$ - as one expects for a single-band quasi-2D metal with unitary curvature[50] - with a $k$-dependent anisotropy that is washed out at either end of the studied temperature range as well as in the high-field limit.

## Failiure of Boltzmann formalism in overdoped cuprates

This evolution of $\ell(\phi,T,H)$ is also expected to determine the form of the in-plane MR $\Delta\rho_{ab}(T)$. Intuitively, for the same FS geometry, $\Delta\rho_{ab}(T)$ should be comparatively small at both ends of the measured temperature range. At high-$T$, $\Delta\rho_{ab}(T)$ will be reduced due to the smallness of $\omega_c\tau$. At low-$T$, the effective isotropy of $\ell(\phi)$, coupled with the weak anisotropy in $k_F(\phi)$, will lead to near cancellation of the magnetoconductance $\Delta\sigma_{xy}/\sigma_{xx}$ by the (square of the) Hall angle $\sigma_{xy}/\sigma_{xx}$.[51] At low but finite temperatures, $\Delta\rho_{ab}(T,H)$ will grow in magnitude (provided that the anisotropy in $\ell(\phi)$ grows at a faster rate than its orbitally-averaged length diminishes). At sufficiently high field strengths, however, $\Delta\rho_{ab}(T,H)$ will always saturate, as the anisotropy in $\ell(\phi)$ is ultimately averaged out.

The SCTIF captures all of these essential features. Extended Data Figure 4a shows the simulated field-dependence of the in-plane MR for OD Tl2201 with a FS volume corresponding to $p = 0.28$ using the same (ADMR-derived) parameters listed in Extended Data Table 3. The in-plane response has been modelled using the following SCTIF-based expression:

$$\sigma_{ij} = \frac{e^3 B}{2\pi^2 \hbar^2 c} \int_0^{2\pi} d\phi \int_0^\infty d\phi' \frac{v_i(\phi) v_j(\phi-\phi')}{\omega_c(\phi)\omega_c(\phi-\phi')} exp\left(\int_\phi^{\phi'} \phi''/\omega_c(\phi'')\tau(\phi'') d\phi''\right) \qquad (2)$$

where $v_x = k_F(\phi)\omega_c \cos(\phi - \zeta) \hbar/(eB)$ and $v_y = k_F(\phi)\omega_c \sin(\phi - \zeta) \hbar/(eB)$ and $\zeta$ is the angle between $k_F$ and $v_F$. That is, anisotropy in $v_F$ follows from anisotropy in $k_F$ with the assumption that the mass renormalisation is isotropic.

There are a number of notable discrepancies between the measured in-plane MR and that predicted by the SCTIF. Firstly, as illustrated in panel b of Extended Data Figure 4, the absolute magnitude of $\Delta\rho = \rho(\mu_0 H = 35$ T$) - \rho(\mu_0 H = 0)$ within the SCTIF formalism is much smaller than found in experiment (see Figure 2a of the main manuscript for comparison). This discrepancy is most striking at the lowest temperatures where the calculated MR effectively vanishes once the total scattering rate (now dominated by impurity scattering) becomes isotropic in $k$-space whereas the experimental MR remains large. Secondly, the calculated field-dependence at high field strengths is sub-linear, indicating a tendency towards saturation, and not strictly linear as in experiment. This is most evident in the derivative plots shown in Extended Data Fig. 4c. Finally, when plotted vs $H/T$ as done in Extended Data Figure 4d, the quadrature scaling and the collapse of d$\rho$/d$H$ found in experiment are completely absent.[52]

Whilst it is evident that the ADMR-derived parameterisation of $k_F(\phi)$ and $\tau^{-1}(\phi,T)$ is unable to reproduce the quadrature MR reported in this work, it is interesting to explore whether a different parameterisation (i.e. ignoring the ADMR-derived parameterisation because $c$-axis transport is, for some reason, intrinsically different to in-plane transport - see, e.g. ref. [53]) is capable of simultaneously reproducing the MR and Hall responses. The Fermi surface of OD Tl2201 is well established, having been corroborated by ADMR, ARPES, quantum oscillations and DFT calculations. We therefore choose to fix $k_F(\phi)$ and primarily investigate whether different forms of $v_F(\phi)$ and $\tau^{-1}(\phi,T)$ are able to model the in-plane magnetotransport more successfully. To this end, we have performed a number of alternative simulations, summarised in Extended Data Figure 5. In each simulation, the FS parameterisation ($k_F(\phi)$, $v_F(\phi)$) is shown in column 1, the scattering rate ($\tau^{-1}(\phi,T)$) in column 2, the derivative of the MR with and without $H/T$ scaling in columns 3 and 4, respectively, and the Hall coefficient $R_H(H)$ in column 5.

For comparison, the top row of Extended Data Fig. 5 (simulation 1) uses the ADMR-derived para-meterisation of $\tau^{-1}(\phi,T)$ (column 2) with the minor simplification that there is no residual anisotropy at $T = 0$ K and that the growth in the anisotropic component is strictly $T$-linear. As before, the resulting MR (column 3) is strongly $T$-dependent and an order of magnitude smaller than that seen in experiment. Additionally, there is no extended region of high-field linearity. Instead, any region of $H$-linearity is simply an inflection point resulting from the anisotropy being progressively averaged out at sufficiently high magnetic fields. Quadrature scaling is not observed (column 4) though, in agreement with Section 3.2, $R_H(H,T)$ (column 5) closely matches experiment.

As previously discussed, fitting of the ADMR was not sensitive to anisotropy in $v_F(\phi)$ and $\tau^{-1}(\phi)$ individually, but to the product $v_F\tau$. Tight-binding modelling of ARPES measurements on Tl2201 with $T_c$ = 30 K[37] suggests that $v_F(\phi)$ has an anisotropy of ~ 50%. Furthermore, the anisotropy in $\tau^{-1}(\phi, T = 10 K)$ was found to be inverted (maximal along $(\pi, \pi)$, i.e, the nodal directions) when compared to that derived from ADMR. In simulation 2 (row 2 of Extended Data Figure 5), we take $v_F(\phi)$ from the tight-binding model that was used to model the ARPES data and also incorporate an inverted anisotropy in $\tau^{-1}(\phi)$ at low $T$ such that the mean-free-path ($\ell = v_F\tau$) is roughly isotropic, consistent with the ADMR at low $T$. Also in accordance with ADMR, it is assumed that the anisotropy in $\tau^{-1}(\phi)$ becomes maximal along $(\pi, 0)$ with increasing temperature. Thus, simulation 2 yields a quantitatively similar MR and $R_H$ to that found in simulation 1.

We note that in ref. [37], a factor of ~ 4 anisotropy in the low-$T$ scattering rate was determined (again with maxima where the ADMR predicts minima). If the anisotropy in $v_F\tau(\phi)$ is still required to be consistent with ADMR at high temperatures, the anisotropy in $v_F$ and $\tau$ will necessarily cancel at finite temperature. That is, effective isotropy is achieved not at low-$T$, but at some elevated temperature. Thus, the MR will be finite at low-$T$, zero at intermediate-$T$, and finite at high-$T$. This parameterisation is therefore inconsistent with the experimentally observed MR. It also fails to capture the Hall coefficient as the field-dependence of $R_H$ is predicted to be positive at low $T$ in this scenario, contrary to the experimental observation.[25]

Although there is no extended region of $H$-linearity (and therefore no possibility of quadrature scaling) in either simulation 1 or simulation 2, one might ask what is required for the maximum slope in the $H$-linear MR to be $T$-independent in order to reproduce at least one key feature of the quadrature scaling and with a magnitude that is comparable to experiment. The minimum change in the parameterisation to achieve this is to impose an anisotropy ratio in $\tau^{-1}(\phi)$ that is both large (~ 7) and independent of temperature. This scenario is modelled in simulation 3 (row 3 of Extended Data Fig. 5). The resultant MR has a maximum slope that is indeed independent of temperature and is of the same order of magnitude as seen in experiment (albeit a factor of 2 lower). There is still no extended region of $H$-linearity, however, and importantly, the curves do not collapse when plotted against $H/T$. An important consequence of fixing the anisotropy ratio is that the low-field $R_H$ also becomes $T$-independent. Thus, it is not possible to reconcile a $T$-independent maximum slope in the MR with a highly $T$-dependent $R_H$ simply by changing the magnitude of the anisotropy in $\tau^{-1}(\phi)$.

Recently, Grissonnanche et al. used the SCTIF to model ADMR measurements in La$_{1.6-x}$Nd$_{0.4}$Sr$_x$CuO$_4$ (Nd-LSCO) with a Sr concentration $x$ = 0.24 i.e. near the end of the pseudogap phase.[54] There, the ADMR-derived parameterisation was able to reproduce the observed in-plane MR showing a $H^2$ to $H$-linear crossover. In Nd-LSCO at this doping level, the FS undergoes a Lifshitz transition as the Fermi level crosses through the vHs near the zone boundary. Proximity to the saddle point near $(\pi, 0)$ generates regions of positive and negative curvature on the FS as well as an in-plane anisotropy in $v_F(\phi)$ that is approximately one order of magnitude larger than that of Tl2201.[37,38] In addition, the impurity scattering rate $1/\tau_0$ is also found to be highly anisotropic, while the $T$-dependent ($T$-linear) component of the scattering rate is isotropic.[54] In light of these findings, we have considered whether such a parameterisation of $\tau^{-1}$ can also be applied to Tl2201. It is important to note that the large $v_F$ (and $\tau^{-1}$) anisotropy used in the modelling of Nd-LSCO arises from proximity of the Fermi level to a vHs at the zone edge. Band structure calculations of Tl2201 (supported by ADMR and quantum oscillation studies) indicate that the zone edge lies far below the vHs. Because of the lack of evidence for such a large anisotropy in $v_F$ in Tl2201 (from ADMR, APRES or DFT), this has not been incorporated into our simulations.

In simulation 4 (row 4 of Extended Data Fig. 5), we consider a form of $\tau^{-1}(\phi)$ that is highly peaked, similar to the one presented in ref. [54], with the proviso that the $T$-dependence of the mean scattering rate maintains the correct form of $\rho(T)$. Intriguingly, this highly peaked form of the scattering rate shifts the $H$-linear inflection point to higher fields making $\Delta\rho(H)$ resemble the high-field linearity over the field-range measured in this work. However, regardless of the precise parameters used, the combination of a $T$-independent anisotropic term and a growing isotropic term will always result in an effective anisotropy ratio that decreases as $T$ increases. This makes the maximum d$\rho$/d$H$ decrease with increasing $T$ and as such, quadrature scaling can never be achieved. In addition, $R_H(T)$ also decreases with $T$ (the inverse of what is seen in experiment). It therefore appears that this parameterisation, whilst possibly relevant to Nd-LSCO, cannot reproduce the in-plane transport properties of Tl2201.

The final simulation (simulation 5) in Extended Data Figure 5 considers the case of OD Bi2201. While the FS of OD Bi2201 has not been probed as extensively by multiple experiments as that of Tl2201, it is clear that it has essentially the same geometry within the strange metal regime. The primary differences between Tl2201 and Bi2201 are the substantial (factor of 3) anisotropy in $v_F(\phi)$[55] (presumed to arise from the closer proximity of the vHs in the latter) and a factor of 10 increase in $\rho_0$ (and thus $1/\tau_0$). Simulation 5 (row 5 of Extended Data Fig. 5) thus uses the Tl2201 FS but includes a form of $v_F(\phi)$ and $\tau(\phi)$ appropriate for OD Bi2201.[55] As in simulation 2, the $\pi/4$ shift between the anisotropy in $v_F(\phi)$ and $\tau_0^{-1}(\phi)$ results in a partial cancellation of the total effective anisotropy of the system. Both the magnitude of the MR and the field-sensitivity of $R_H$ are reduced accordingly. As in the case of Tl2201, in order to rectify the situation, one would have to impose a very large $T$-independent anisotropy in $\tau^{-1}(\phi)$. In passing, we also note here that the $H$-linear MR (and associated $H/T$ scaling) is seen to be a robust feature of not only Tl2201 and Bi2201, but several strange metals too, including the multi-band pnictides[10] and chalcogenides,[13] as well as a number of other cuprate families with very different FS shapes and anisotropies. This universality, coupled with the extreme fine tuning required to model such behaviour within the SCTIF, suggests strongly that its origin lies beyond conventional Boltzmann analysis.

On more general grounds, combining the residual resistivity of OD Bi2201 $\rho_0 \sim 100$ $\mu\Omega$cm and the in-plane Hall coefficient $R_H(0) \sim 1 \times 10^{-9}$ m$^3$/C,[25] we obtain an estimate for $\omega_c\tau$ of order $R_H/\rho_0 \sim 10^{-3}$/T. This in turn provides an estimate for the strength of the orbital magnetoconductance $\Delta\sigma_{ab}/\sigma_{ab}$ of $(\omega_c\tau)^2 \sim 1 \times 10^{-6}$/T$^2$. The magnitude of the quadratic MR found in our OD Bi2201 crystals at low fields is almost 3 orders of magnitude larger than this Drude estimate. While a large anisotropy can increase the size of the MR, the previous Boltzmann analysis shows that even in-plane anisotropies in $\omega_c\tau \sim 3$-4 generate a MR that is still more than an order of magnitude smaller than found experimentally in Bi2201 and that varies quadratically with field up to 30 T. Hence, just as in Tl2201, the MR observed in OD Bi2201 cannot be attributed easily to conventional orbital effects.

Finally, we turn to address the observation of quantum oscillations in OD Tl2201 in the presence of a putative incoherent sector. We first note that despite concerted experimental effort over a few decades, quantum oscillations (QOs) have only been observed to date in Tl2201 with doping levels $p > 0.275$.[48] The observation of QOs in both the far OD regime of Tl2201 and in several cuprates (YBa$_2$Cu$_3$O$_{7-\delta}$,[56] YBa$_2$Cu$_4$O$_8$[57] and HgBa$_2$CuO$_{4+\delta}$[58]) in the underdoped regime, could be taken as evidence that there are coherent quasiparticles which can complete cyclotron orbits at these compositions. This does not however rule out the co-existence of an incoherent sector. In the underdoped case, ARPES measurements suggest that incoherence exists around the anti-nodal directions in $k$-space and QOs are possible because of a reconstruction of the Fermi surface, probably resulting from a charge density wave, which connects the nodal arcs together. In the far OD regime however, the QO frequency suggest the whole (1+$p$) Fermi surface is traversed. Any $k$-space regions of incoherence would have to be very small (a few percent of the total circumference of the orbit) for magnetic breakdown to be effective in overcoming this at, for example, $\mu_0H$ = 20 T. The question then is how could such small regions of incoherence give rise to an MR which exceeds the SCTIF calculations by more than one order of magnitude?

We do not claim to have the answer to this question, or the similar one of why the SCTIF explains so well the $c$-axis MR but not the in-plane MR, but it is worth noting that QOs are a thermodynamic quantity and can arise in circumstances far departed from a conventional Fermi liquid state. For example, it has been shown theoretically that QOs can exist even in band-insulators.[59] Moreover, QOs have been observed in the Kondo insulator SmB$_6$[60,61] and more recently in YbB$_{12}$.[62] QOs have also been observed in the superconducting state where the coherent quasiparticles are separated in real space from the gapped regions (See ref. [63] for a review). There have also been theories of QOs in strongly interacting quantum critical non-Fermi liquids.[64] QOs therefore do not necessarily indicate a conventional metallic state. However, it remains a theoretical challenge to explain their existence inside the strange metallic state of OD cuprates.

## Kohler's rule vs. quadrature MR in overdoped cuprates

For normal metals with a quadratic low-field MR, Kohler's rule states that $\Delta\rho/\rho(0,T) \propto (H/\rho(0,T))^2$. Hence for a metal obeying Kohler's rule, all the magnetoresistance curves collapse once plotted versus $(H/\rho(0,T))^2$. To illustrate Kohler's rule violation in OD cuprates, we have plotted $\Delta\rho_{ab}/\rho_{ab}(0,T)$ versus $(H/\rho_{ab}(0,T))^2$ for a Bi2201 sample with $T_c$ = 13 K in

Extended Data Fig. 6a. A second way of illustrating Kohler behaviour is to plot the product $\Delta\rho_{ab}\cdot\rho_{ab}(0,T)$ (normalised to $\Delta\rho_{ab}(H)$ at 1 T). If Kohler scaling is obeyed, this quantity should be independent of temperature. However, as demonstrated in Extended Data Fig. 6c, $\Delta\rho_{ab}\cdot\rho_{ab}(0,T)$ exhibits a marked $T$-dependence, appearing to diverge as $T \rightarrow 0$. (Here, fits were made to the low field part of the MR with the quadratic function $f(x) = A(\mu_0 H)^2$.)

Below $p^*$, the MR response in high-$T_c$ cuprates appears to become dominated by an orbital contribution, that either obeys[65] conventional Kohler's scaling (in the underdoped regime) or violates it (at optimal doping).[66] In both cases, the $T$-dependence of $\Delta\rho_{ab}/\rho_{ab}(0,T) = 1/(A + BT^2)^2$ links the MR to the square of the Hall angle (though in UD cuprates, the longitudinal resistivity also approaches more of a Fermi-liquid response).[67,68]

For a long time, it has been believed, rightly or wrongly, that the in-plane MR of overdoped cuprates could also be captured by the so-called modified Kohler's rule in which $\Delta\rho_{ab}/\rho_{ab}(0,T) = A \cot(\vartheta_H)^{-2}$. In this case, $\Delta\rho_{ab}\cdot\rho_{ab}(0,T) = a(\rho_{ab}\tan(\vartheta_H))^{-2} \sim a(\rho_{xy})^2$, i.e. $\Delta\rho_{ab}\cdot\rho_{ab}(0,T)$ should have the same $T$-dependence as the square of the Hall coefficient $R_H$. As shown in Extended Data Fig. 6b, while this relation appears to work well at elevated temperatures, it clearly breaks down below around 60 K where $\Delta\rho_{ab}\cdot\rho_{ab}(0,T)$ grows while $R_H^2$ becomes smaller. One possible origin of this enhancement in $\Delta\rho_{ab}\cdot\rho_{ab}(0,T)$ is the onset of paraconductivity contributions. As shown in Extended Data Fig. 6d, however, superconducting fluctuations only manifest themselves - as an upturn in the derivative of $\rho_{ab}(0,T)$ - below about 30 K.

In a system governed by quadrature MR, the $T$-dependence of the quadratic $A$ term must vary as $1/T$. This is obvious from the fact that the derivatives of the field dependence scale once plotted versus $H/T$. Hence, for the quadrature MR to be realized, the product $A\cdot T$ should be constant. As shown in Extended Data Fig. 6e, this relation does indeed appear to hold over a decade in temperature. We therefore conclude that the $T$-dependence of the low-field $H^2$ MR is better described by quadrature MR than by the modified Kohler's rule. The fact that at high temperatures, the quadratic $A$ term has a $T$-dependence that follows expectations for both the modified Kohler's rule and the quadrature MR may explain why previous low-field measurements[47] could not make this distinction.

## Determination of $\beta$ in Tl2201 and Bi2201

The quadrature form of $\rho_{ab}(H,T)$ is characterised by a crossover from $H$-linear MR at high $H/T$ to a $H^2$ dependence at low $H/T$. The crossover scale $(\mu_0 H/T)^*$ is universal for all $T$ as illustrated by the collapse of the derivatives of the field sweeps taken at different temperatures onto a single curve (Figure 2 of the main manuscript) and is determined by the ratio $\beta$ of the quadrature field and temperature coefficients. As it happens, $(\mu_0 H/T)^*$ does not depend on the absolute value of the measured $\rho_{ab}(H,T)$ and is therefore insensitive to errors arising from the determination of the sample dimensions, for example. It is therefore possible to make a direct comparison between different samples, as done in panels k and l of Figure 2.

Since the crossover scale $(\mu_0 H/T)^*$ is larger for Bi2201 than for Tl2201 a different approach was used to obtain $\beta$ for Bi2201. There, a fit was made to the derivatives plotted against $H/T$ (for example, see panel d of Figure 2 in the main manuscript), where only the normal state data was used in the fit. This approach allows for all the normal state data to be fitted at once and gives a representative value of $\beta$ for Bi2201.

As mentioned above, in certain Tl2201 samples, hydrostatic pressure was also applied to enable the evolution of $\beta$ to be studied as a function of $T_c$ independent of the level of disorder. Moreover, whereas previously it was only possible to directly compare $\beta$ ($\sim \gamma/\alpha$), by changing pressure on a single sample, it became possible to study the relative changes in $\alpha$ and $\gamma$ individually with $T_c$. In one sample, $T_c$ was suppressed in small steps from 35 K to 26 K with the application of 2 GPa. These results are presented elsewhere.[69]

## Planckian dissipation and Zeeman coupling

The specific form of the quadrature scaling discovered by Hayes and co-workers in P-Ba122[10] implies that magnetic field and temperature influence the resistivity through the transport relaxation rate in a similar manner. Moreover, since the ansatz can be written as:

$$\rho(H,T) - \rho_0 = \alpha k_B T \sqrt{1 + (\gamma \mu_B \mu_0 H / \alpha k_B T)^2} \qquad (3)$$

the residual resistivity $\rho_0$ has no influence on the MR scaling. This is in marked contrast to what is observed in normal metals where $\Delta\rho/\rho(0)$ scales with $(\omega_c\tau_{tr})^2$, where $1/\tau_{tr} = 1/\tau_0 + 1/\tau_{in}$, $1/\tau_0$ is the impurity (elastic) scattering rate and $1/\tau_{in}$ is the $T$-dependent (inelastic) scattering rate. The independence of the quadrature MR from the strength of the impurity scattering (or more precisely, from $\rho_0$) has been confirmed not only here in Tl2201 and Bi2201 -- despite almost one order of magnitude difference in their respective $\rho_0$ - but also previously in FeSe$_{1-x}$S$_x$ (near the nematic QCP) where the quadrature MR was found to be comparable in samples with $\rho_0$ values that differed by a similar factor of around 5.[13] Crucially, observance of the quadrature scaling also demands that the zero-field resistivity associated with the MR response has a $T$-linear dependence. If we rewrite $\sqrt{1+(\gamma\mu_B\mu_0H/\alpha k_BT)^2}$ as $\sqrt{1+(\gamma^*\omega_\hbar\tau_\hbar)^2}$, we may define an effective $\omega_\hbar\tau_\hbar$ product for the Planckian sector where $\omega_\hbar = e\mu_0H/m^*$ ($m^*$ here is the corresponding 'cyclotron' mass) and $\tau_\hbar = \hbar/\alpha k_BT$. Moreover, since $\mu_B = e\hbar/2m_e$, we find that $\gamma^* = \alpha\gamma/2\alpha(m^*/m_e)$. Assuming $\gamma^* \sim 1$ and $m^* = m_e$, we obtain $\alpha \sim 0.3$ for Tl2201 and $\alpha \sim \pi$ for Bi2201. Thus, one can generate directly from the quadrature MR response a relaxation rate compatible with Planckian dissipation without a priori knowledge of parameters such as (a) the carrier density or (b) quasiparticle mass that (a) would determine the absolute magnitude of the resistivity and (b) may be irrelevant for a description based on Planckian dissipation. Nevertheless, notions of cyclotron frequency and mass do not sit comfortably with the fact that the MR response appears to be non-orbital in nature (as deduced from the lack of angle dependence in the in-plane MR of both Bi2201 and Tl2201).

In light of this and the fact that the zero-field resistivity is associated with Planckian dissipation with temperature as the only relevant energy scale, it is also feasible that the $H/T$ scaling of the MR originates from the spin sector, with the magnetic field introducing a second energy scale via Zeeman coupling. Even in this overdoped regime, the system may still behave like a doped Mott-insulator where a separate spin system may be identifiable. The antiferromagnet at half filling is very isotropic and the angular momentum associated with the charged currents may dissipate by coupling to the non-conserved spin angular momentum. The difficulty is however that for this to happen a strong spin-orbit coupling $\lambda$ is required. In the cuprates $\lambda$ should be small, as confirmed by the isotropy of the spin system of the Mott insulator. A potential loophole may relate to the unanticipated strong spin-orbital locking of the quasiparticles observed recently by spin-resolved ARPES.[70]

In P-Ba122[10] and FeSe$_{1-x}$S$_x$,[13] the quadrature MR term is much reduced when $H$ is rotated into the conducting plane. In order to account for such anisotropy, one would require the Zeeman term to be highly anisotropic. In elemental bismuth, such anisotropy (in the hole band) was shown to arise from the combined effects of spin-orbit coupling and multiple bands.[71] FeSe$_{1-x}$S$_x$ and P-Ba122 are, of course, both multi-band (semi-)metals and the pivotal role of spin-orbit coupling in creating large spin-space anisotropy in FeSe$_{1-x}$S$_x$ has already been discussed.[72] Tl2201 and Bi2201 are single-band cuprates, while spin-orbit coupling is thought to play only a minor role. The near-isotropy of the MR response in these OD cuprates may therefore support a picture in which magnetic field influences the incoherent sector by adding a second energy scale that leads to a $H$-linear growth in the resistivity. Such a scenario might then account for the difference in the anisotropy in the MR response in both the copper-based and iron-based superconductors.

Finally, while the near-isotropy of the MR response for in-plane currents points more towards a spin-based, rather than orbitally driven, origin for the $H$-linear MR, we note here that a similar isotropic component to the $c$-axis MR is not evident in any of the ADMR measurements performed to date. However, it is important to recognise that while spin effects might generate an isotropic contribution for a particular current direction with respect to field orientation, it is not necessarily the case that they will generate isotropic contribution with respect to current direction, particularly if the processes for charge transport in the two orthogonal directions are different. The in-plane MR that we observe is significantly larger than what we would expect from the analysis of the $c$-axis MR, suggesting that it may be an additional contribution. Alternatively, one might need to go beyond the relaxation time approximation and consider vertex corrections for in-plane transport, corrections that do not necessarily play a role in $c$-axis transport. It has been argued,[73,74] for example, that for layered materials, $c$-axis transport can be described as a product of single-particle in-plane spectral functions and thus can be computed without vertex corrections. Hence, even though ADMR may be observed,[74] it does not imply that the transport processes in the two orthogonal directions are identical.

## Methods references

**Data availability:** The data that support the plots within this paper and other findings of this study are available from the Bristol data repository, data.bris, at https://doi.org/XXXX

**Extended Data Figure captions**

**Extended Data Fig. 1 | Zero-field resistivities of Tl2201 and Bi2201.** Zero field, ambient pressure resistivity $\rho_{ab}(T)$ curves for representative **a**, Tl2201 and **b**, Bi2201 crystals investigated in this study. Note the super-linear $T$-dependence for all samples. The spread in absolute magnitudes of $\rho_{ab}(T)$ is higher in the Tl2201 crystals due to the fact that they were mounted for pressure measurements and as such, their absolute resistivities were harder to quantify accurately.

**Extended Data Fig. 2 | Quadrature scaling in overdoped Tl2201. a,** $\rho_{ab}(H,T)$ as measured in Tl2201 with $T_c$ = 26.5 K. $H^2$ behaviour cedes to a $H$-linear resistivity at high fields. **b, c,** Scaling plots of $(\rho_{ab}(H,T) - \rho_{ab}(0,0))/T$ vs $H/T$ for OD Tl2201 ($T_c$ = 26.5 K). As shown in panel **c**, there is a clear break down of the scaling at low $H/T$. **d, e,** Scaling plots of $(\rho_{ab}(H,T) - \rho_{ab}(0,T))/T$ vs $H/T$ for the same sample where $\rho(0,T) = \mathcal{F}(T) = \rho_0 + A_gT + BT^2$. Note that $A_g$ does not correspond to $A$, the full $T$-linear coefficient of the zero-field resistivity, since part of that is contained within the quadrature form. The inclusion of these additional $T$-dependent terms makes the data collapse over the full $T$-range. Taking the derivative with respect to $H$ (as done in the main text) provides another means of isolating the quadrature MR from $\mathcal{F}(T)$. The dashed lines in all panels represent the quadrature expression $\Delta\rho_{ab}(H) = \alpha k_B T\sqrt{1 + (\beta\mu_0 H/T)^2}$ ($\rho_0$ = 15.5 μΩcm, $A_g$ = 0.14 μΩcm/K, $B$ = 0.003 μΩcm/K$^2$, $\alpha k_B$ = 0.04 μΩcm/K, $\gamma\mu_B$ = 0.20 μΩcm/T). **f,** The derivatives with respect to magnetic field of the measured curves shown in **a**. **g,** When plotted against $H/T$, the derivatives presented in **f** collapse onto a universal curve (bar those sections of each field sweep that are in the mixed state).

**Extended Data Fig. 3 | Success of ADMR-derived modelling of the in-plane transport of OD Tl2201. a,** The $c$-axis ADMR of Tl2201 with $T_c$ = 15 K measured at 50 K and at various (labelled) azimuthal angles taken from ref. [48]. **b,** Projection of the in-plane FS derived from the ADMR fitting. **c,** Schematic showing the isotropic $T^2$ component (black solid line) and anisotropic $T + T^2$ component (red solid line) of the scattering rate as deduced from the ADMR fitting. **d,** Black dots: $\rho_{ab}(T)$ data for OD Tl2201 ($T_c$ = 15 K) in which superconductivity has been suppressed by a magnetic field (**H** || $c$)[46] and corresponding simulation based on the ADMR fitting.[48] The difference in the residual resistivities is likely due to the fact that different samples have been used in the two studies.[46,48] **e,** Corresponding simulation for $R_H(T)$.[48] **f,** Simulation of $R_H(H) = \rho_{xy}(H)/H$ at various temperatures as indicated. **g,** Same simulation data plotted versus $H/\rho(0)$ where here, $\rho(0)$ is the zero-field resistivity at each temperature. **h,** $R_H(H)$ versus $H/\rho(0)$

data taken from ref. [25]. For OD Tl2201 ($T_c$ = 25 K) for comparison with the simulation in panel **g**. The larger absolute values of $R_H$ in panel **h** relative to panel **g** are due to the fact that the high-field data in **h** are taken on a sample with a higher $T_c$ value where the anisotropy in $\tau^{-1}(\phi)$ is expected to be larger.

**Extended Data Fig. 4 | Failure of ADMR-derived modelling to reproduce quadrature scaling. a,** The field dependence of the longitudinal resistivity ($\rho(T)$) determined with the SCTIF using parameters derived from the ADMR parameterization for OD Tl2201. **b,** $\Delta\rho$ – the change in $\rho_{xx}$ with field – at selected temperatures. **c,** Corresponding derivative plots of $\rho(H)$ showing distinctly non-quadrature behaviour. **d,** As a consequence, the data fail to collapse when plotted against $\mu_0H/T$.

**Extended Data Fig. 5 | Failure of the SCTIF to reproduce both the MR and Hall response.** Simulations of the MR and Hall responses within the SCTIF given different parameterisations of $v_F(\phi)$ and $\tau^{-1}(\phi, T)$. In each simulation, the experimentally determines FS ($k_F(\phi)$) has been used. Note that the SCTIF is slow to converge at low fields and so the simulations do not extend all the way to $H$ = 0. **Simulation 1,** ADMR-derived parameterisation of OD Tl2201 albeit with no anisotropy at $T$ = 0 and an anisotropic term that increases strictly linearly with $T$. **Simulation 2,** A scenario incorporating the $v_F$ anisotropy derived from tight-binding modelling of ARPES measurements.[37] **Simulation 3,** A scenario in which the anisotropy ratio of $\tau^{-1}(\phi)$ is strictly $T$-independent in order to generate an MR with a maximum slope that is also independent of temperature (reminiscent of quadrature scaling). **Simulation 4,** A scenario in which a similar $\tau^{-1}(\phi)$ parameterisation to the one used to model Nd-LSCO[54] is applied to OD Tl2201. **Simulation 5,** Simulation for OD Bi2201 with an enhanced anisotropy in $v_F$ and $\tau^{-1}(T = 0)$ consistent with ARPES.[51] **Column 1,** The FS parameterisations $k_F(\phi)$ and $v_F(\phi)$. **Column 2,** $\tau^{-1}(\phi, T)$. **Column 3,** $d\rho/d(\mu_0H)$ versus $H$. **Column 4,** $d\rho/d(\mu_0H)$ versus $H/T$. **Column 5,** $R_H(H, T)$.

**Extended Data Fig. 6 | Kohler versus quadrature scaling in Bi2201. a,** $\Delta\rho_{ab}/\rho_{ab}(0)$ plotted versus $(H/\rho_{ab}(0))^2$ for a Bi2201 sample with $T_c$ = 13 K. In a system that shows Kohler scaling, these curves would collapse. Clearly, that is not the case here. **b,** $\Delta\rho_{ab}$ plotted versus $H^2$ for the same Bi2201 sample. The dotted lines are fits to the function $f(x) = A(\mu_0H)^2$ in the regions where the MR is strictly quadratic. Note that the quadrature form of the MR is only purely quadratic in the zero field limit while fits to the data are taken at finite field ranges. Our simulations have shown that fitting up to $\mu_0H_{max} = \beta \cdot T$ (with $\beta$ as given in Fig. 2 of the main text) agrees with the zero field limit within a few percent and falls within our experimental error. **c,** $T$-dependence of $A \cdot \rho_{ab}(0)$ (with $A$ taken from the fits in panel **b**) compared to the square of the Hall coefficient $R_H^2$. **d,** Temperature derivative of $\rho_{ab}(T)$ for the same sample. Note that the onset of superconducting fluctuations appears only below 30 K. **e,** The product $A \cdot T$ plotted over the full temperature range. The dotted line is a guide to the eye.

**Extended Data Table 1. | Tl2201 samples studied.** The $T_c$ values are defined as the temperature below which $\rho_{ab}(T)$ falls below the noise floor. The doping levels are determined from the $T_c$ values, as explained in the text.

**Extended Data Table 2. | Bi2201 samples studied.** The $T_c$ values are defined as the temperature below which the zero-field resistivity falls below the noise floor. The doping levels are determined from the $T_c$ values, as explained in the text. Bi2201 samples labelled with a $T_c$ < 1 K were measured down to 1.4 K, and although they show a clear onset of superconductivity, they do not become fully superconducting. We have therefore specified their doping level as $p$ = 0.27.

**Extended Data Table 3. | ADMR simulation parameters.** Parameters derived from ADMR fits used for the SCTIF calculations for a $T_c$ = 15 K sample of Tl2201.[48]

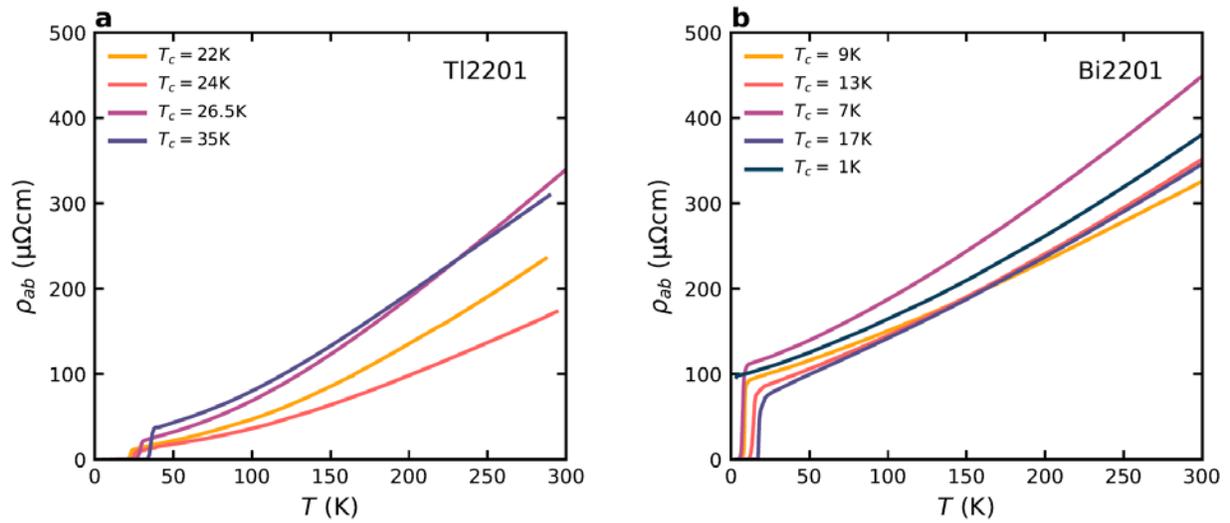

**Extended Data Figure 1**

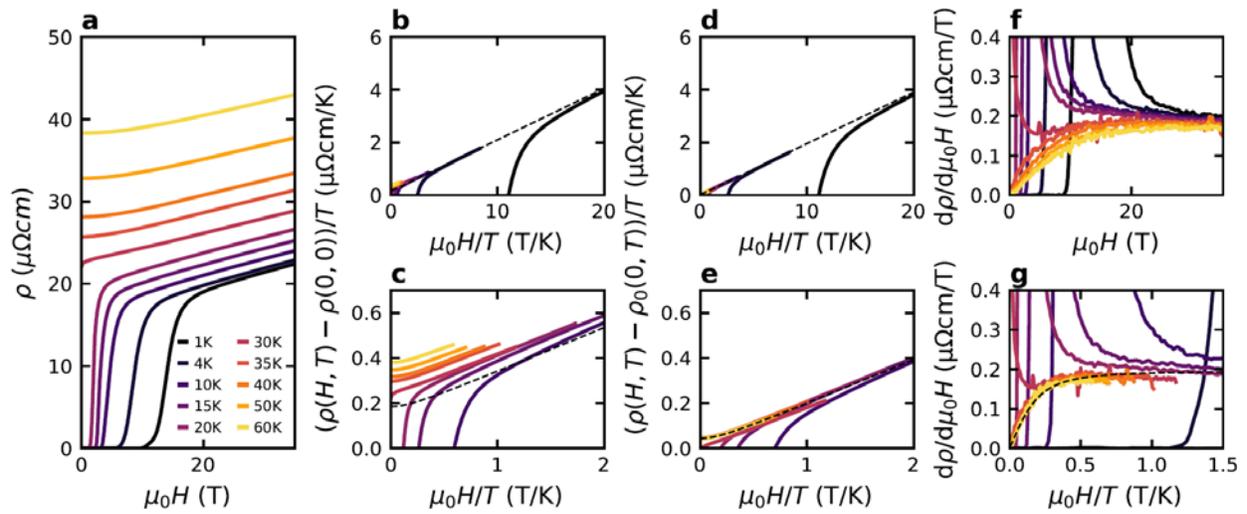

**Extended Data Figure 2**

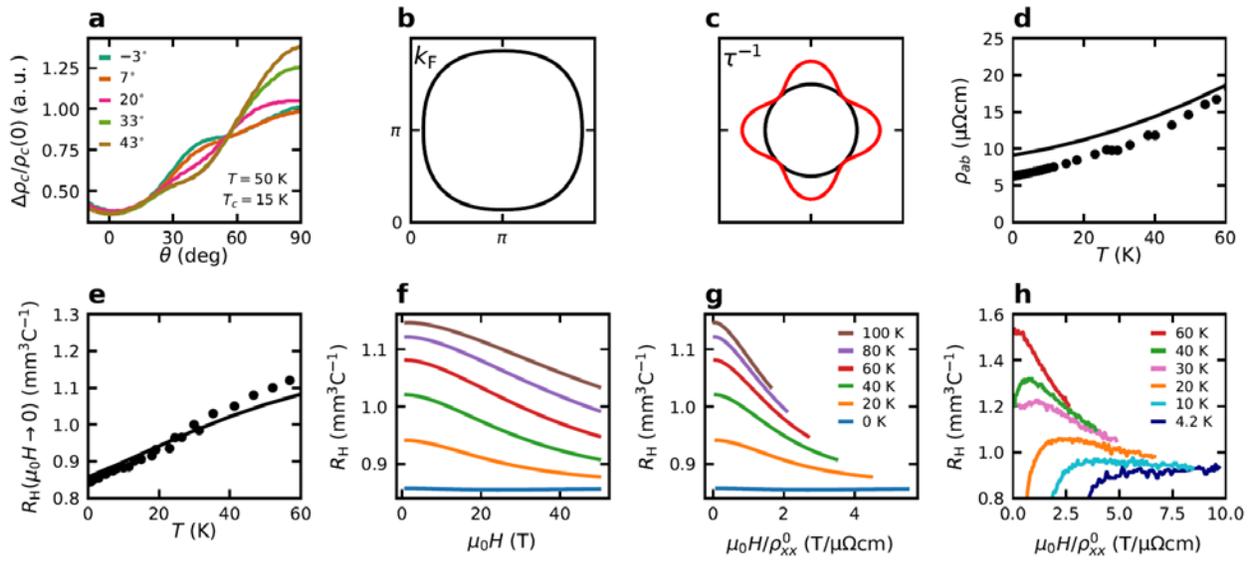

**Extended Data Figure 3**

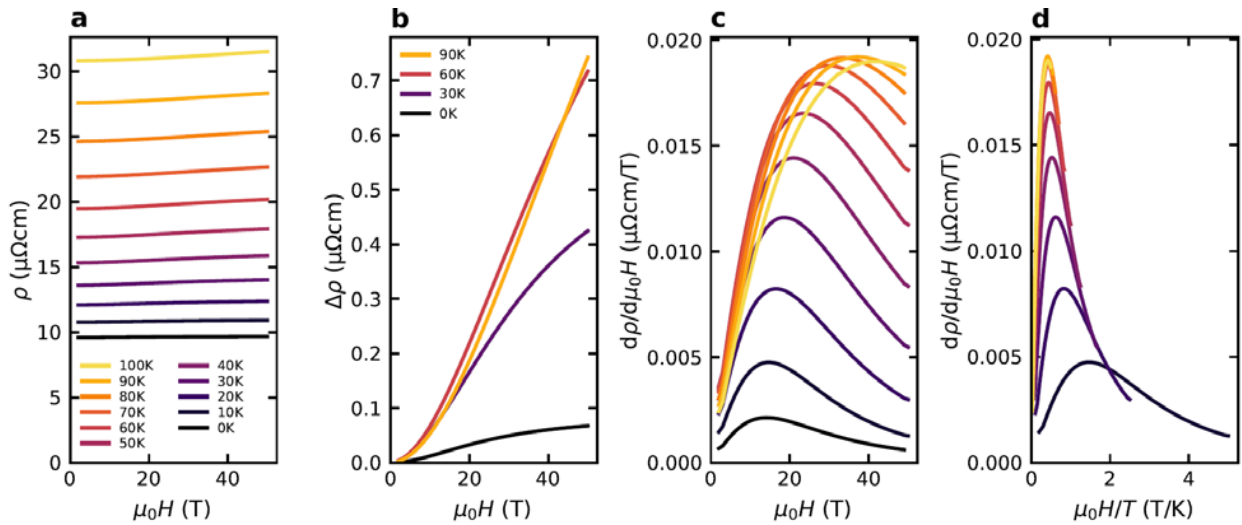

**Extended Data Figure 4**

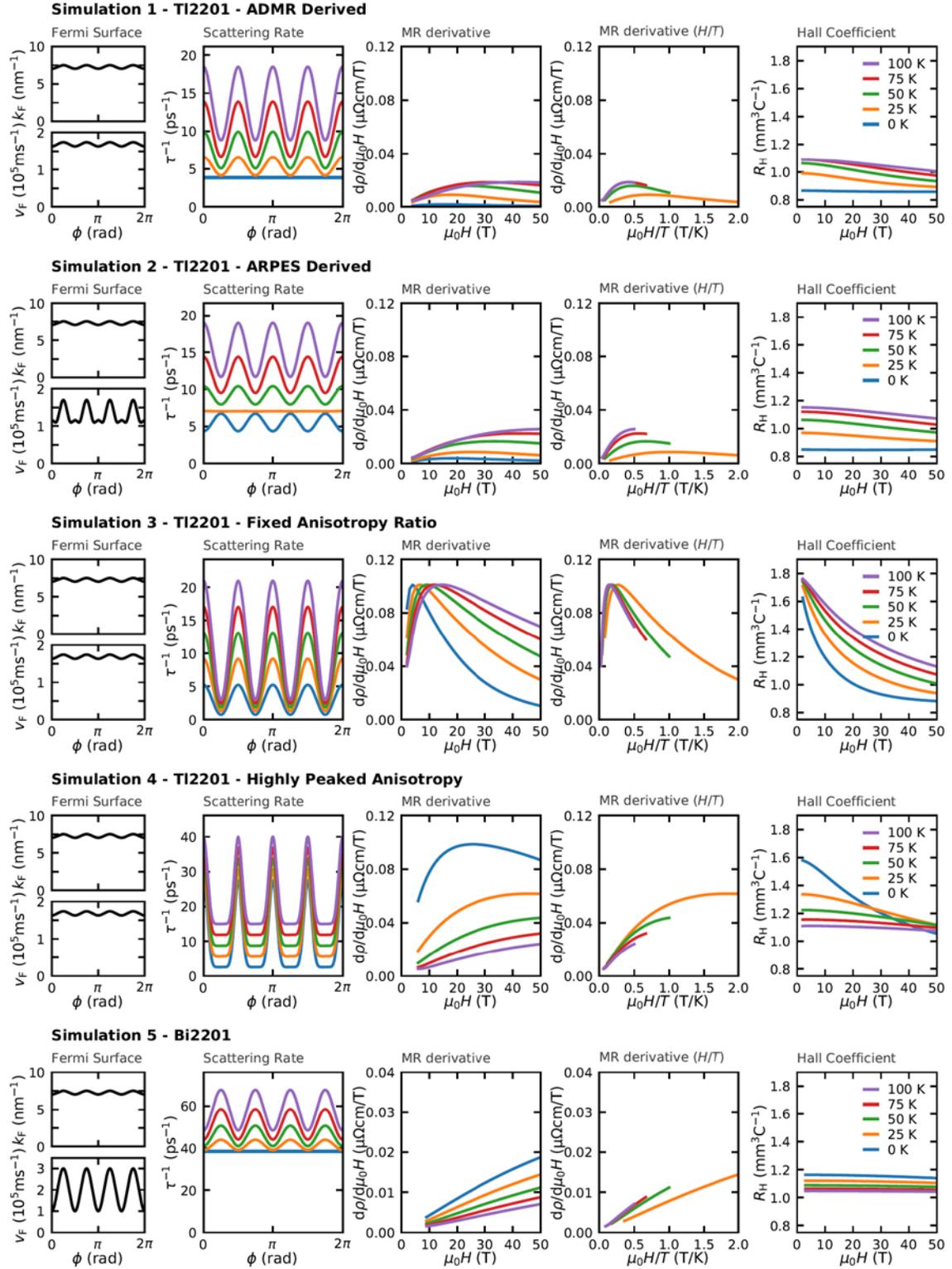

**Extended Data Figure 5**

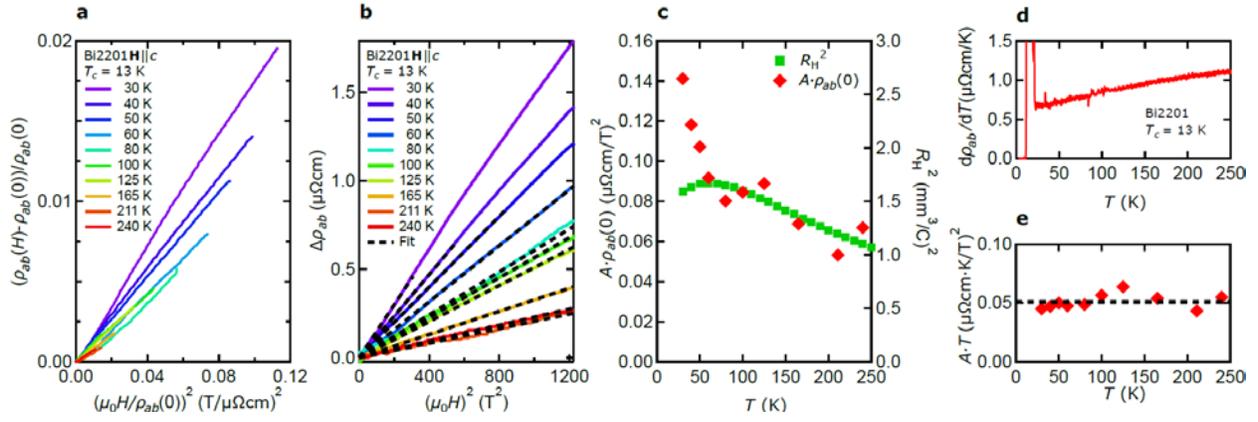

**Extended Data Figure 6**

| Tl2201 Sample | Inferred Doping | $T_c$(0 GPa) | Field Orientations | Pressures |
|---|---|---|---|---|
| #1 | 0.283 | 20 K<br>14 K (2 GPa) | **H** ∥ c | 0 – 2 GPa |
| #2 | 0.280 | 22 K | **H** ∥ c | 0 GPa |
| #3 | 0.280 | 22 K | **H** ∥ ab, **H** ∥ c | 0 GPa |
| #4 | 0.279 | 23 K | **H** ∥ c | 0 GPa |
| #5 | 0.277 | 24 K | **H** ∥ ab, **H** ∥ c | 0 GPa |
| #6 | 0.274 | 26.5 K | **H** ∥ ab, **H** ∥ c | 0 GPa |
| #7 | 0.262 | 35 K | **H** ∥ ab (0 GPa), **H** ∥ c | 0 – 2 GPa |
| #8 | 0.256 | 40 K<br>26 K (2 GPa) | **H** ∥ c | 0 – 2 GPa |

**Extended Data Table 1**

| Bi2201 Sample | Inferred Doping | $T_c$ | Field Orientations |
|---|---|---|---|
| #1 | 0.27 | < 1 K | **H** ∥ *c* |
| #2 | 0.27 | < 1 K | **H** ∥ *ab*, **H** ∥ *c* |
| #3 | 0.27 | < 1 K | **H** ∥ *ab*, **H** ∥ *c* |
| #4 | 0.27 | < 1 K | **H** ∥ *ab*, **H** ∥ *c* |
| #5 | 0.258 | 7 K | **H** ∥ *ab*, **H** ∥ *c* |
| #6 | 0.255 | 9 K | **H** ∥ *ab*, **H** ∥ *c* |
| #7 | 0.247 | 13 K | **H** ∥ *ab*, **H** ∥ *c* |
| #8 | 0.239 | 17 K | **H** ∥ *c* |
| #9 | 0.239 | 17 K | **H** ∥ *ab*, **H** ∥ *c* |

**Extended Data Table 2**

| Parameter | Value |
|---|---|
| $k_{00}$ | 0.0728 nm$^{-1}$ |
| $k_{40}/k_{00}$ | -0.033 |
| $A_{iso}$ | 2.41 |
| $B_{iso}$ | 2.82 x 10$^{-2}$ K$^{-2}$ |
| $A_{ani}$ | -9.94 x 10$^{-2}$ |
| $B_{ani}$ | 4.15 x 10$^{-2}$ K$^{-1}$ |
| $C_{ani}$ | 3.29 x 10$^{-4}$ K$^{-2}$ |

**Extended Data Table 3**